\documentstyle[preprint,aps]{revtex}

\begin{document}

\draft
\title{\bf Electron-electron interactions in
one- and three-dimensional mesoscopic disordered rings: a
perturbative approach}

\author{Michel Ramin, Bertrand Reulet and H\'el\`ene Bouchiat}
\address{Laboratoire de Physique des Solides, associ\'e au
CNRS\\
B\^at. 510, Universit\'e  Paris--Sud, 91405, Orsay , France.}

\maketitle

\begin{abstract}
{We have computed persistent currents in a
disordered mesoscopic ring in the presence of small
electron-electron interactions, treated in first order
perturbation theory. We have investigated both a contact
(Hubbard) and a nearest neighbour interaction in 1D and 3D. Our results
show that a repulsive Hubbard interaction produces
a paramagnetic contribution to the average current (whatever
the dimension) and increases the value of the typical
current. On the other hand, a nearest neighbour repulsive interaction
results in a diamagnetic contribution
in 1D and paramagnetic one in 3D, and tends to decrease the value
of the typical current in any dimension. Our study is based
on numerical simulations on the Anderson model and is
justified analytically in the presence of very weak
disorder. We have also investigated the influence of the amount of disorder
and of the statistical (canonical or grand-canonical) ensemble.}
\end{abstract}
\newpage

Since the experimental discovery \cite{levy90,chan91,mail93}
of persistent currents in phase-coherent
mesoscopic rings pierced by a magnetic flux, a great deal of
theoretical work has been devoted to the understanding of the
order of magnitude of these currents, which is difficult to explain
within a single electron theory.  Both the
amplitude of the average current
measured in a many-rings experiment\cite{levy90},
and the typical current measured in single
ring experiments \cite{chan91,mail93}
are found to be,
at least in the
diffusive regime, between 1 and 2 orders of magnitude larger than
their expectations\cite{mont90,alts91,schm91,oppe91,bouc91,cheu89l}.
In order to explain this dicrepancy between
experiment and theory,
several authors have studied the influence of electron-electron (e-e)
interactions on the persistent currents.
Since the work of Altshuler and Aronov\cite{alts85b} based on diagrammatic
theory, it has been indeed known that transport properties of quantum coherent
disordered
systems are affected by e-e interactions. Inspired by these results,
Ambegaokar and Eckern\cite{ambe90} have shown that
the presence of short-range interactions  gives rise to a contribution to
the average persistent current, whose sign depends on the sign of
the interactions. The low flux current is found to be
paramagnetic for repulsive and diamagnetic for attractive interactions.
It has also been suggested that the typical value of the current could
be strongly enhanced by interactions and disorder independent
\cite{ecke92,mull93}. In
contrast with these results which are valid for 3D disordered rings in the
diffusive regime, numerical 1D exact diagonalisations
have found that the current is decreased in the
presence of repulsive nearest neighbour interactions
\cite{bouz94}. The case of long
range Coulomb interactions has also been investigated\cite{abra93,kato94},
with the result
that the current is decreased for a very weakly disordered ring
and increased at higher disorder. However
from these studies performed in 1D it is very difficult
to anticipate the behavior of a realistic 3D
multichannel ring, which is very difficult to handle exactly numerically.

These considerations have motivated the present work where
we have studied the influence of a small e-e interactions, on
disordered rings in 1D and 3D. We have compared the cases of a
pure contact Hubbard and a first neighbour interaction. The
e-e interaction is treated within first order of perturbation
on the Anderson model as follows:
\begin{equation}
{\cal H}=\sum_{i,j}\left[t_{ij} c^+_{i}c_{j}+c.c.\right] +
\sum_{i}w_{i}c^+_ic_i
\end{equation}
with
$t_{ij}=\exp\left(2i\pi\Phi/\Phi_0(x_{i}-x_{j})/L_{x}\right)$,
where $i$ and $j$ are nearest neighbours, and where $w_i$ is the onsite
disorder potential uniformly distributed in the interval $[-W/2,W/2]$.
This hamiltonian is exactly diagonalized in the presence of disorder .

We first consider the one-band Hubbard model,
characteristic of contact interactions described by the
hamiltonian
$\displaystyle{\cal H}_{int}=U\sum_{i}n_{i\uparrow}n_{i\downarrow}$
with $n_{i\sigma}=c^+_{i\sigma}c_{i\sigma}$ and $U>0$ for repulsive
interactions. The first order
correction to the energy of a N electrons state reads :
\begin{equation}
\left\{
\begin{array}{lcll}
\delta E(2p)&=&\displaystyle\sum_{\alpha,\beta=1}^p{\cal U}(\alpha,\beta)
&{\rm if\ }N=2p\\
\delta E(2p+1)&=&\displaystyle\delta E(2p)+
\sum_{\alpha=1}^{p}{\cal U}(\alpha,p+1)&{\rm if\ }N=2p+1
\end{array}\right.
\end{equation}
Here ${\cal U}(\alpha,\beta)$ describes the interaction
energy between electrons in the states $\alpha $ and $\beta$
which are single electron eigenstates of the Anderson
hamiltonian.
\begin{equation}
{\cal U}(\alpha,\beta)=U\sum_i |\psi_\alpha(i)|^2|\psi_\beta(i)|^2
\end{equation}
where $\psi_\alpha(i)$ denotes the value of the one-electron wavefunction
at the site $i$.
We can also write the correction to
the ground state with $2p$ electrons in the form:
\begin{equation}
\delta E(2p)=U\sum_i n_p^2(i)
\label{arga}
\end{equation}
where  $\displaystyle n_p(i)=\sum_{\alpha=1}^p|\psi_\alpha(i)|^2$ is the
electronic density at the site $i$ with the first $p$
eigenstates occupied by only one electron. Expression (\ref{arga}) is very
similar to the result obtained by Argaman and Imry\cite{arga93}.
We deduce the correction to the current given by:
\begin{equation}
\delta I(N)=-\frac{\partial { \delta
E(N)}}{\partial { \Phi}}
\end{equation}
and compute the canonical (C) and grand-canonical (GC)
average of this current in 3D
in the diffusive regime (the averaging is done over the filling, between
$1/4$ and $3/4$ of the spectrum).
In the GC case, only configurations with an even number of electrons
contribute. In fig. \ref{fig1}a these quantities are
compared to the average C and GC
currents without interactions. We can see that,
unlike the non-interacting case, both
statistics give identical results for $\langle\delta I\rangle$
which has the same paramagnetic sign (in zero flux) as
the canonically average current in the absence of interactions.
Furthermore, the effect of
interactions is to enhance the current, which is in agreement
with analytical results\cite{ambe90}.
 From the first
order correction to the energy , it is also possible to
compute the quantity
$\langle\delta (I^2(N))\rangle=2\langle I(N)\delta I(N)\rangle$
which is the first order correction in U to the typical
current $\langle I^2\rangle$. We observe in fig. \ref{fig1}b
that $\langle I\delta I\rangle$, in both C and GC statistics,
 is always positive and of the order of
$\sqrt{\langle I^2\rangle\langle\delta I^2\rangle}$, which means that
there is a strong correlation between the current of $N$ electrons and the
correction to this current for the same number of electrons.
We have made this calculation for different
values of the disorder. The first harmonics of average and
typical currents are presented in fig. \ref{fig2}
as a function of the disorder $W$. We observe that $\langle I_2\rangle$,
$\langle\delta I_2\rangle$, $\sqrt{\langle I^2_1\rangle}$ and
$\sqrt{\langle I_1\delta I_1\rangle}$ decrease as $W^{-2}
\propto l_{e}$, the mean free path.
This dependence of $\langle\delta I_2\rangle$ with the disorder is in
complete agreement with analytical calculations\cite{ambe90}. Nevertheless,
as already pointed out\cite{bouc91}, the disorder dependence of the C average
$\langle I_2\rangle$ for this small ring disagrees with analytical
predictions\cite{alts91,schm91,oppe91}.
In conclusion we have found that the
effect of a contact interaction results in an increase of
both average and typical currents. In the following we
demonstrate this result analytically in the weak disorder
limit. Eigenstates of the Anderson hamiltonian in 1D without
disorder are plane waves of the type
$\psi_{n,\Phi}(x)=\exp(i2\pi(n-\Phi/\Phi_0)x/L)/\sqrt L$. In the presence
of a very weak disorder, the degeneracy occuring at
$\Phi=0$ and $\Phi=\Phi_0/2$ is lifted,
resulting in the following wavefunctions:
\begin{equation}
\left\{
\begin{array}{rcl}
\psi_{n,\Phi}^{+}(x)&=&\sqrt{2/L}\cos[2\pi(n-{\Phi/\Phi_0})x/L+\chi_n]\\
\psi_{n,\Phi}^{-}(x)&=&\sqrt{2/L}\sin[2\pi(n-{\Phi/\Phi_0})x/L+\chi_n]\\
\end{array}\right.
\end{equation}
where $\chi_n$ is a phase factor which depends on the particular realisation
of the disorder.
We can easily deduce the following value of $\delta E(2p)$~:
\begin{equation}
\delta E(2p) = \left\{
\begin{array}{rl}
(p^2+\frac12)/L & {\rm for\ } (\Phi=0 {\rm\ and\ } p=2N)
{\rm\ or\ }(\Phi=\Phi_0/2 {\rm\ and\ } p=2N+1)\\
p^2/L& {\rm otherwise}\\
\end{array}\right.
\end{equation}
The resulting $\delta I(\Phi)$ presents an antisymmetric
paramagnetic peak in the vicinity of $\Phi=0$ (resp.
$\Phi_0/2$)  for an even (resp. odd) number of electron
pairs, in agreement with the symmetry property of the 1D
electron spectrum, according to which:
$I(N,\Phi) \sim I(N+1,\Phi+\Phi_0/2)$,
valid in the limit $N \gg 1$. This behavior gives rise
to a current contribution $\delta I$ which has the same sign as
in the absence of interactions (i.e.
paramagnetic in zero flux for $2p$  electrons) in agreement
with our numerical findings (see fig. \ref{fig3}a).

In comparison with the Hubbard model we
also considered the short-range nearest neighbours
interactions studied by Bouzerar {\it et al.}\cite{bouz94}.
The corresponding hamiltonian reads~:
$\displaystyle{\cal H}_{int}=U\sum_{i}c^+_ic_ic^+_{i+1}c_{i+1}$.
Two terms appear now in the first order correction to the energy~:
a direct and an exchange term, so that the final correction reads~:
\begin{equation}
{\delta E(2p)}=U\left(2\sum_{i}n_p(i)n_p(i+1)-
\sum_{i}\sum_{\alpha,\beta=1}^p\psi_{\alpha}^{*}(i)\psi_{\beta}(i)
\psi_{\alpha}(i+1)\psi_{\beta}^{*}(i+1)\right)
\end{equation}
Note that unlike the case of ref.\cite{bouz94}, here spin has
been taken into account. However we have found that the
exchange term is always much smaller than the direct one,
which means that the expression obtained for spinless
electrons would only differ by a factor two from what we have
calculated.

In 1D, we find a current contribution which, for each
value of the number of electron pairs (which varies from
$\frac14L$ to $\frac34L$, $L$ being the length of the ring)
is opposite in
sign to the current in the absence of interaction,
in agreement with ref.\cite{bouz94}. The
resulting average current gives rise to a diamagnetic
contribution (see fig. \ref{fig3}b).

	On the other hand in 3D, the same type of interaction
gives rise to an average paramagnetic contribution to the
current, which is shown in fig. \ref{fig4}a
for different values of the disorder. We observe like previously
that the current does not
depend on the statistics considered, but its
amplitude is much larger than  the value
obtained for a contact interaction.
The dependence
of the average current on disorder is also very weak in contrast with
the results obtained for the Hubbard interaction (see fig. \ref{fig2}a).
If we now study the results obtained for the corrections to the typical
current in 3D we see that for most values of the flux and for ballistic and
diffusive regimes $\langle I\delta I\rangle$ is negative for the
C statistics
except in the vicinity of $\Phi=0$ and $\Phi=\Phi_0/2$ where it is
slightly positive. In the GC ensemble $\langle I\delta I\rangle$ is always
negative (see for instance fig. \ref{fig4} obtained in the
diffusive regime). The
typical current is thus diminished in the presence of first
neighbour interactions both in 1D and 3D. The variation of the first
harmonics versus disorder are presented in fig. \ref{fig2}.
We observe that $\langle\delta I_2\rangle$
generally does not vary very much with
disorder.

We have also performed an analytical computation of $\delta E(2p)$
(as previously discussed in the Hubbard case) for a 1D ring in the limit of
very weak disorder, and obtained for the first order energy correction~:
\begin{equation}
\delta E(2p) =\left\{
\begin{array}{rl}
[f_1(p)+f_2(p)]/L &
{\rm for\ } (\Phi=0 {\rm\ and\ } p=2N)
{\rm\ or\ }(\Phi=\Phi_0/2 {\rm\ and\ } p=2N+1)\\
f_1(p)/L& {\rm otherwise}\\
\end{array}\right.
\end{equation}
with
\begin{equation}
f_1(p)=2p^2-\left(\frac{\sin p\pi/L}{\sin\pi/L}\right)^2
{\rm\ and\ }f_2(p)=\frac12\cos2p\pi/L
\end{equation}
Note the strong dependence of $\delta E(p)$ on the filling~:
the sign of $\cos2p\pi/L$ determines the flux dependence
of $\delta E(2p)$ and consequently the sign of
$\delta I(N,\Phi)$. When $p$ varies between $\frac14L$ and
$\frac34L$, $\cos2p\pi/L$ is negative. It
results from this that $\delta I(\Phi)$ presents an
antisymmetric diamagnetic peak near $\Phi=0$
($\Phi_0/2$) for an even (odd) number of electron pairs.
In the weak disorder limit this calculation can be generalised to a
multichannel ring. In that case we can separate the contribution of the
interactions to the persistent current into its longitudinal and transverse
part. The longitudinal component can, depending on the filling, lead to
diamagnetism; however, the transverse part of the interaction always gives
rise to paramagnetism, since it is identical to the Hubbard contribution.
This explains why the total contribution is
paramagnetic. Nevertheless, note that the case of anisotropic interactions
may lead to the opposite conclusion.
So we have obtained a confirmation of our numerical results
which show that a first neighbour interaction gives a
diamagnetic contribution to the average current in 1D, and
also tends to decrease the value of the typical current. Only
this last result remains valid in 3D, where the average
current is paramagnetic.

 From our perturbative study of the influence of e-e
interactions on the persitent current in a disordered ring, we
have shown that a repulsive Hubbard interaction enhances the
typical value of the persistent current in any dimension. It also
contributes to the average current with a paramagnetic sign in zero
field. These extra both typical and average contributions to the current
decrease with disorder just like the single electron
quantities. There is thus no evidence that the Hubbard interaction could
attenuate the sensitivity of the currents upon disorder.

A nearest neighbour interaction, on the other hand, tends to
decrease the value of the typical current, while the sign of the average
contribution depends however on the dimensionality of the ring~: it is
diamagnetic in 1D and paramagnetic in a multichannel ring, with very
small disorder dependence. These differences between Hubbard and
nearest neighbour interactions in the one dimensional limit\cite{giamxx}
have also been found analytically.

This work has strongly benefited from the help
and suggestions of A. Altland, Y. Gefen, T. Giamarchi and G. Montambaux.
We thank also J. A. Cowen for having revisited the manuscript.
Numerical simulations have been performed using CRAY
facilities at IDRIS (Orsay). This work was partly supported by grant
from DRET No. 92/181.

\begin{figure}
\caption{
(a): flux dependence of C and GC corrections to the average current,
compared to their values without interactions.
(b): C and GC corrections to the typical current compared to the
typical C current without interactions.
All curves are calculated for a $30\times4\times4$ ring,
for a contact (Hubbard) interaction  with $W=2$ and $U=1$.}
\label{fig1}
\end{figure}

\begin{figure}
\caption{
Dependence of different harmonics of the correction to
average or typical current on disorder. The effect of
Hubbard (subscript "H") or nearest neighbour (subscript "NN") interactions
are compared with the average and typical currents without interaction
(all of them for a $30\times4\times4$ ring with $U=1$).
(a): second harmonics of the average current with Hubbard, nearest neighbour
or no interaction.
(b): first harmonics of the typical current with or without Hubbard
interactions.
(c): first harmonics of the typical current with or without
nearest neighbour interactions.}
\label{fig2}
\end{figure}

\begin{figure}
\caption{
Flux dependence of $\delta I$ for a Hubbard (a) or nearest neighbour (b)
interactions for all
the even fillings of a 1D ring of length $L=16$, from
$\frac14L$ to $\frac 34L$. The circles correspond to the
average within the same range}
\label{fig3}
\end{figure}

\begin{figure}
\caption{
(a): flux dependence of $\langle\delta I\rangle$ for a
nearest neighbour interaction
for three amplitudes of disorder, in the C ensemble in 3D.
(b): flux dependence of (negative) C and GC corrections to the typical
current compared to the typical C value, for a
$30\times4\times4$ ring with nearest neighbour interactions,
with $W=3$ and $U=1$.}
\label{fig4}
\end{figure}

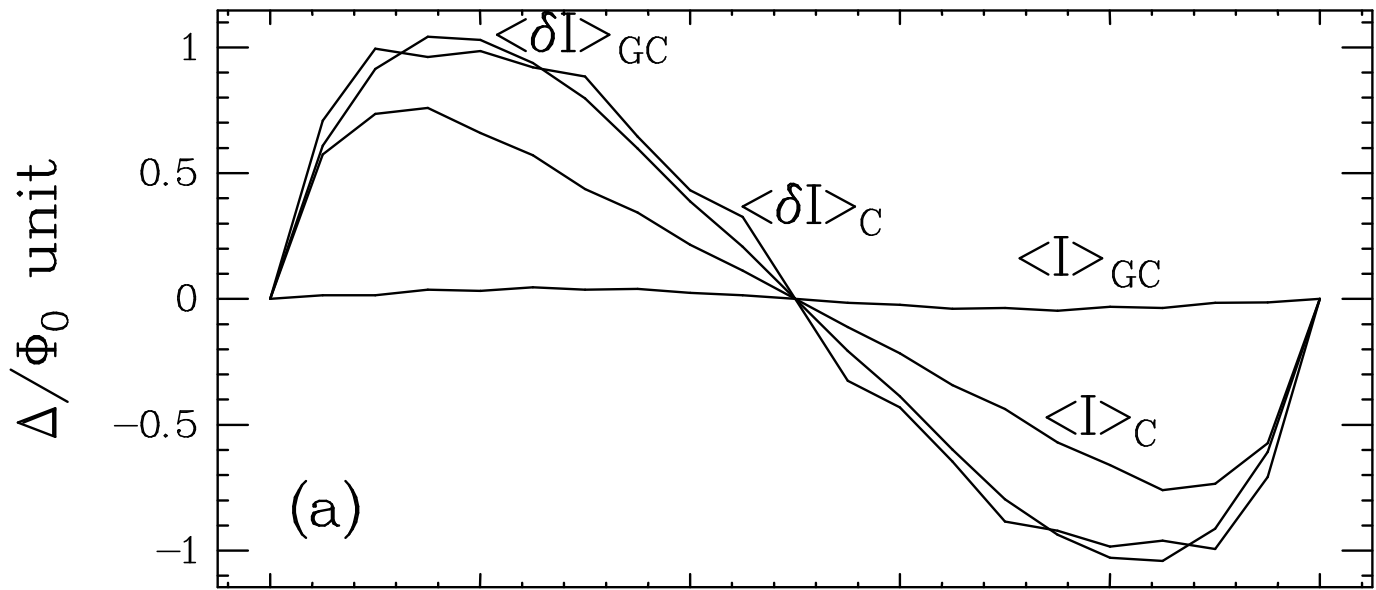
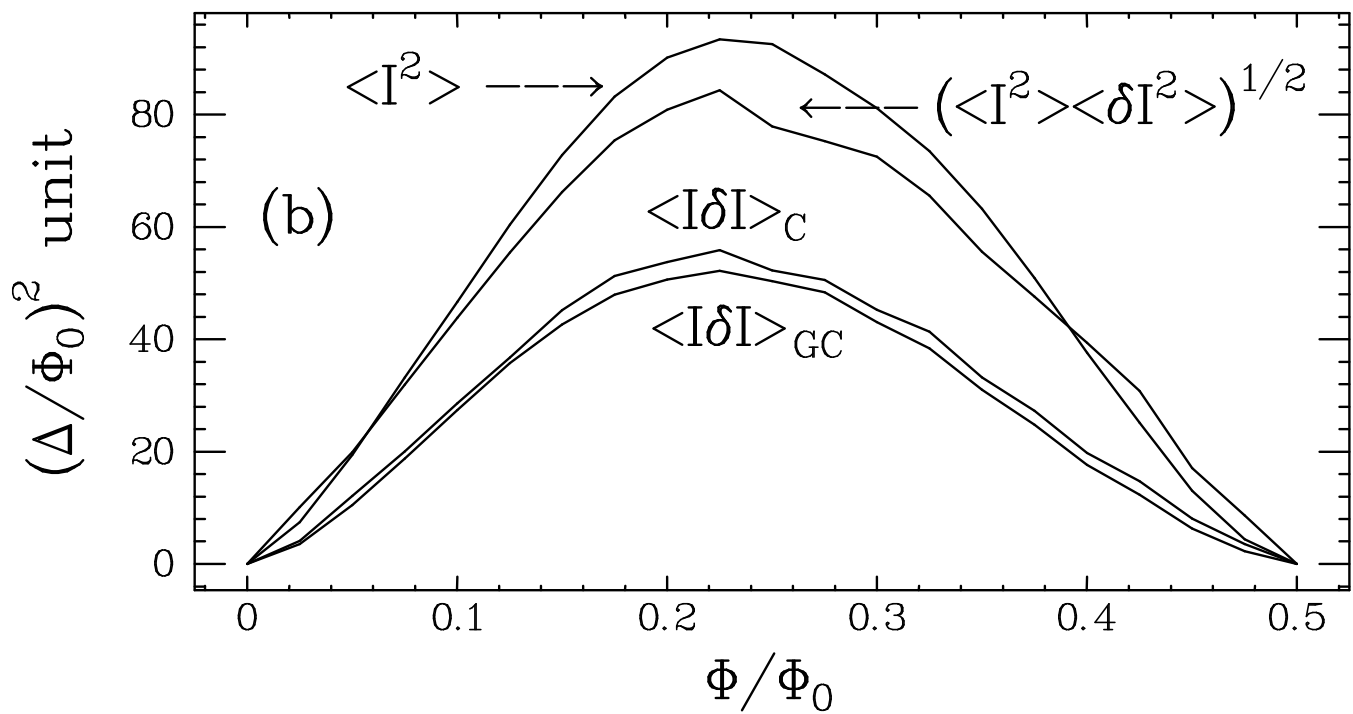
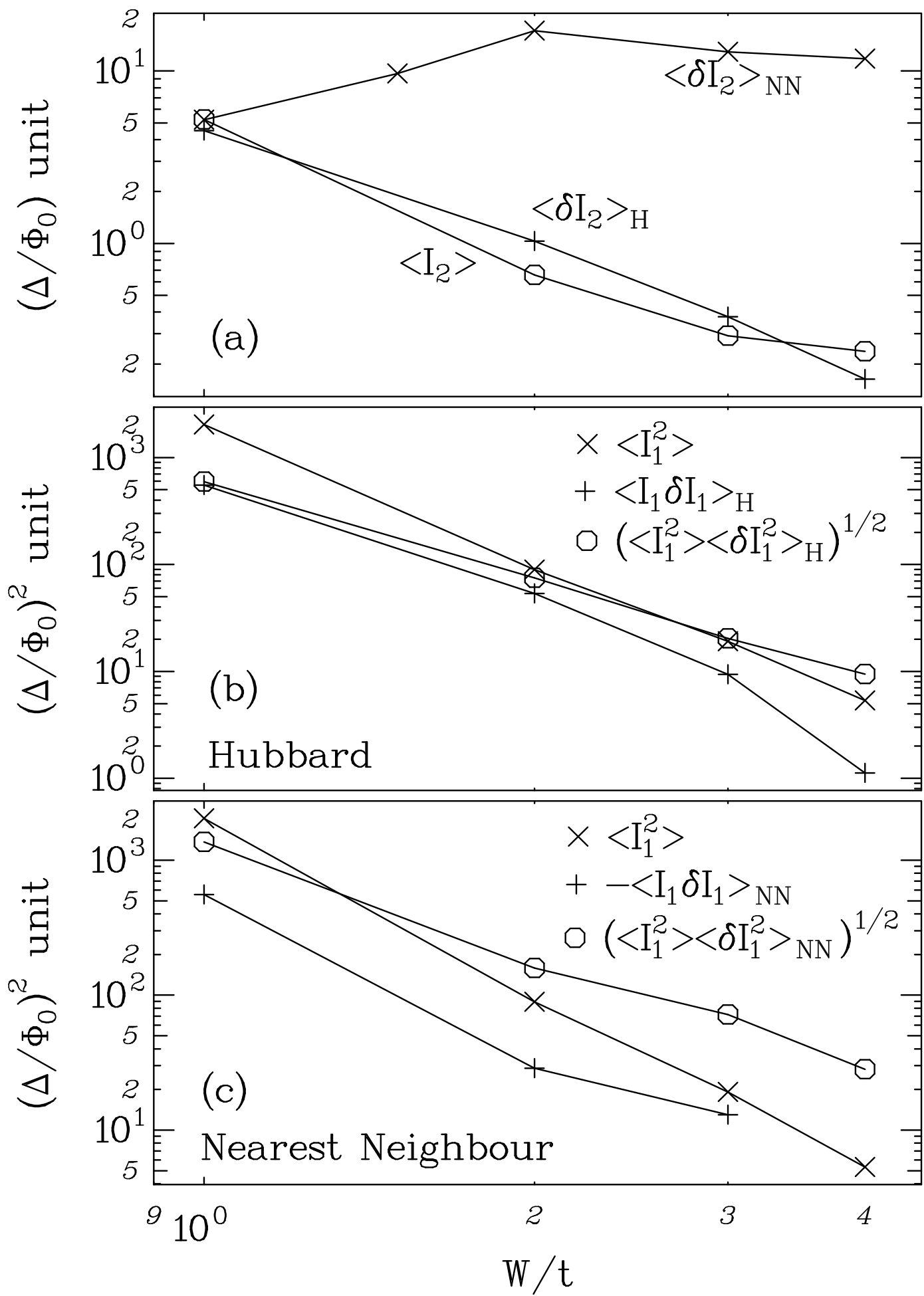
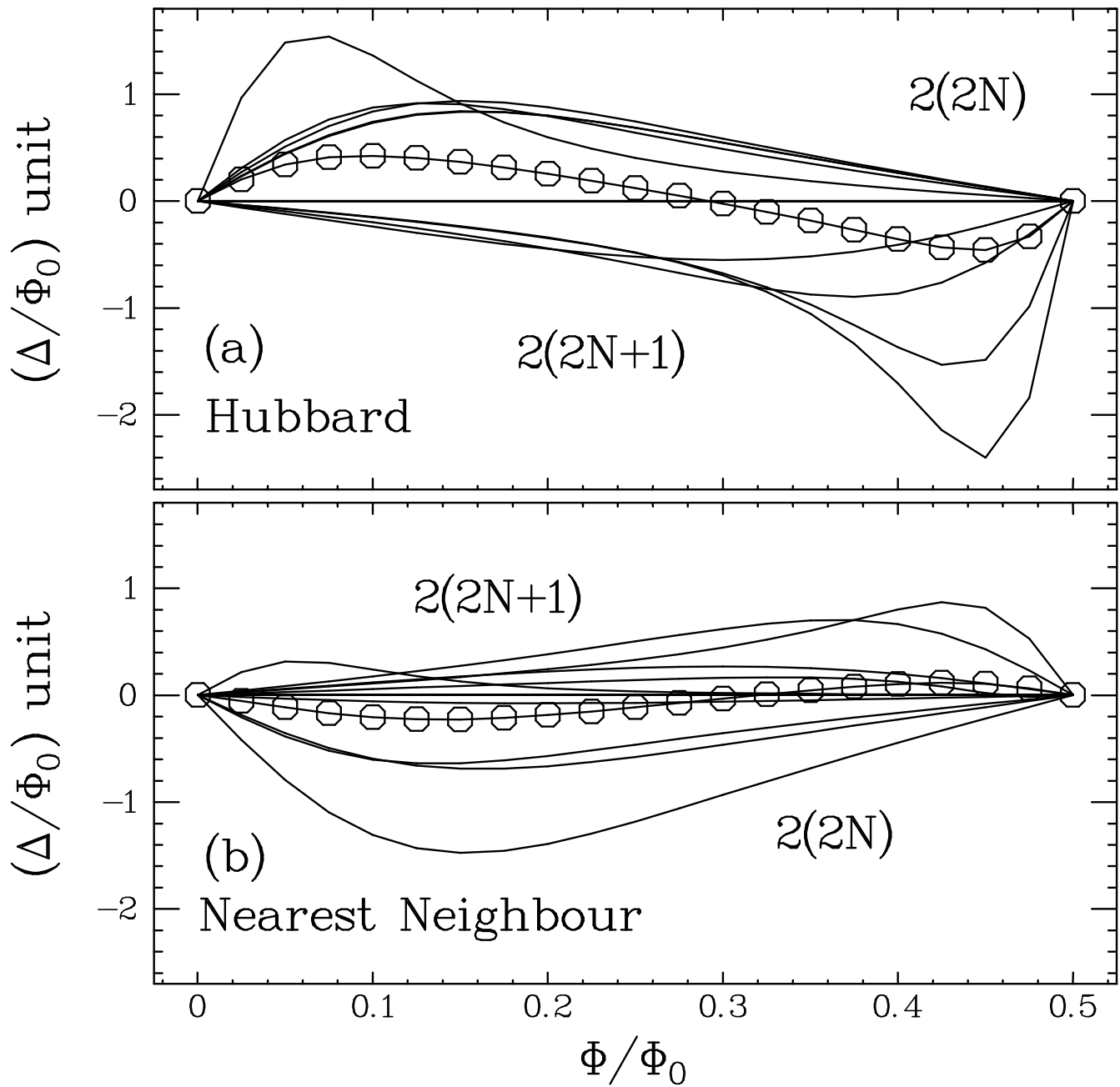
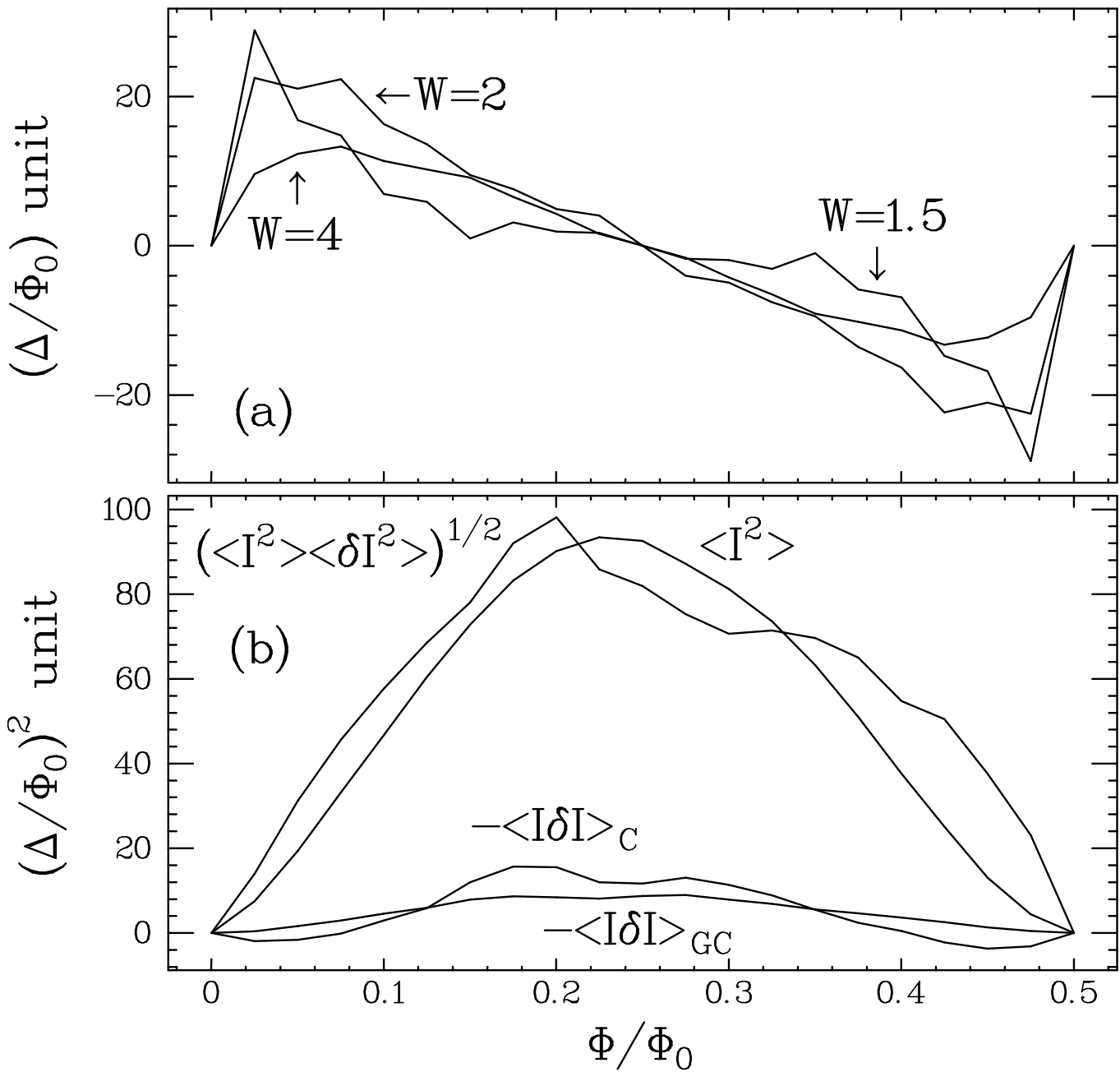
\end{document}